# Knowledge Formation and Inter-Game Transfer With Classical and Quantum Physics

Mads Kock Pedrsen[1], Camilla Clement Borre[1], Andreas Lieberoth[1,2], and Jacob Sherson[1]
[1]Center for Community Driven Research, Aarhus University, Aarhus, Denmark
[2]Danish School of Education, Aarhus University, Aarhus, Denmark
madskock@phys.au.dk
ccborre@hotmail.com
andreas@edu.au.dk
sherson@phys.au.dk

**Abstract:** In order to facilitate an intuitive understanding of classical physics concepts we have developed Potential Penguin – a game where players manipulate the landscape around a sliding penguin in order to control its movement. The learning goal of Potential Penguin is to familiarize players with kinetic energy and potential energy - the energies associated with movement and position in the landscape respectively. The game levels introduce the concepts one by one, as players are tasked with sliding the penguin through a landscape towards a specific location, while keeping the velocity under control. When the player manipulates the landscape, the potential energy of the penguin is changed, which determines the penguin's movement. To build a strong connection between theory and game the analytical expressions for kinetic and potential energy are displayed during play with font sizes continually growing and shrinking according to changes in each energy type.
With Potential Penguin we hope to study whether visualizing the amount of kinetic and potential energy through visible mathematical expressions generates a connection between the intuitive actions taken in the game and the underlying physics concepts.
The knowledge about kinetic and potential energy gained with Potential Penguin can also be used to understand most of the physics behind the citizen science game Quantum Moves, which has the goal of building a working quantum computer. The two games share the principle of the core interaction – manipulating the potential-energy landscape. We aim to investigate whether a proficiency and understanding of Potential Penguin predicts a better performance in Quantum Moves and a deeper understanding of the quantum physics behind that game.

**Keywords:** gamification, citizen science, physics education, inter-game knowledge transfer,

## 1. Introduction

Potential and kinetic energy – the energies associated with position and movement respectively – are fundamental concepts throughout classical physics, since an understanding of the two concepts and their relationship underlies insight into most classical physics phenomena. Traditionally, when high school students are introduced to potential and kinetic energy, they are presented with the analytical expressions and have to work through a couple of simplified problems. In the learning game Potential Penguin students are introduced to potential and kinetic energy and their inter-dependencies though the core mechanics of a puzzle game, which consists of manipulating the potential energy of a sliding penguin in order to steer it to its desired destination while also maintaining control of its velocity. By incorporating the learning goal into the core game mechanics, we expect students to gain an implicit feel for the physics concepts as they learn how to play the game employing their existing sense of intuitive physics (Scherr, Close, & McKagan, 2012; Sherin, 1999). In order to create an explicit connection to classical physics concepts and visualize the more theoretically oriented learning content for the students (Hattie and Yates, 2014) the mathematical expressions for the potential and kinetic energy are displayed with dynamically changing font sizes reflecting the respective amount of energy at any given moment. If embedding the mathematic expressions into the game is effective then other learning games could use the same method to generate explicit connections (Lieberoth and Hansen, 2011; Moon, 2004) between the game mechanics and the learning goals.

The principle of manipulating the potential landscape is shared by the citizen science game *Quantum Moves* (Lieberoth et al, 2014), which has demonstrated (Sørensen et al, 2016) that players given the right spatiotemporal means to manipulate physics problems though a game interface are able to solve extremely abstract, complex and counterintuitive challenges which are otherwise highly intractable. Earlier studies of

Quantum Moves have also demonstrated that the game can be successfully applied to teaching quantum physics (Bjælde, Pedersen, and Sherson, 2014; Magnusen et al, 2014, Pedersen et al, 2016). However, it remains an open question whether an understanding of the physics principles behind the game has a positive impact on the players' game performance. If so it would introduce a new way to enhance the introduction of Quantum Moves such that the game yields better results.

In this paper we propose a study design that will test the following two hypothesises including both Potential Penguin and Quantum Moves:

- Dynamically displaying the mathematical expressions for potential and kinetic energy terms with relative font sizes facilitates explicit understanding of the concepts, and strengthens the tie to concrete analytical problem solving.
- Playing the Potential Penguin game gives an understanding of these physics concepts, which can be transferred to Quantum Moves.

## 2. Game mechanics

The game consists of a penguin sliding in a frictionless topographical landscape of ice hills and valleys (Figure 1). The player can manipulate the height of hills as the penguin is sliding towards a finish line. If the penguin is on the hill when the height is changed, then the potential energy changes as well. As the penguin slides downhill the potential energy is converted into kinetic energy, i.e. it gains speed. The same mechanic applies in reverse when the penguin slides uphill, losing speed. Thus, the students obtain absolute control of the penguin's movements by manipulating the height of the hills in the right way.

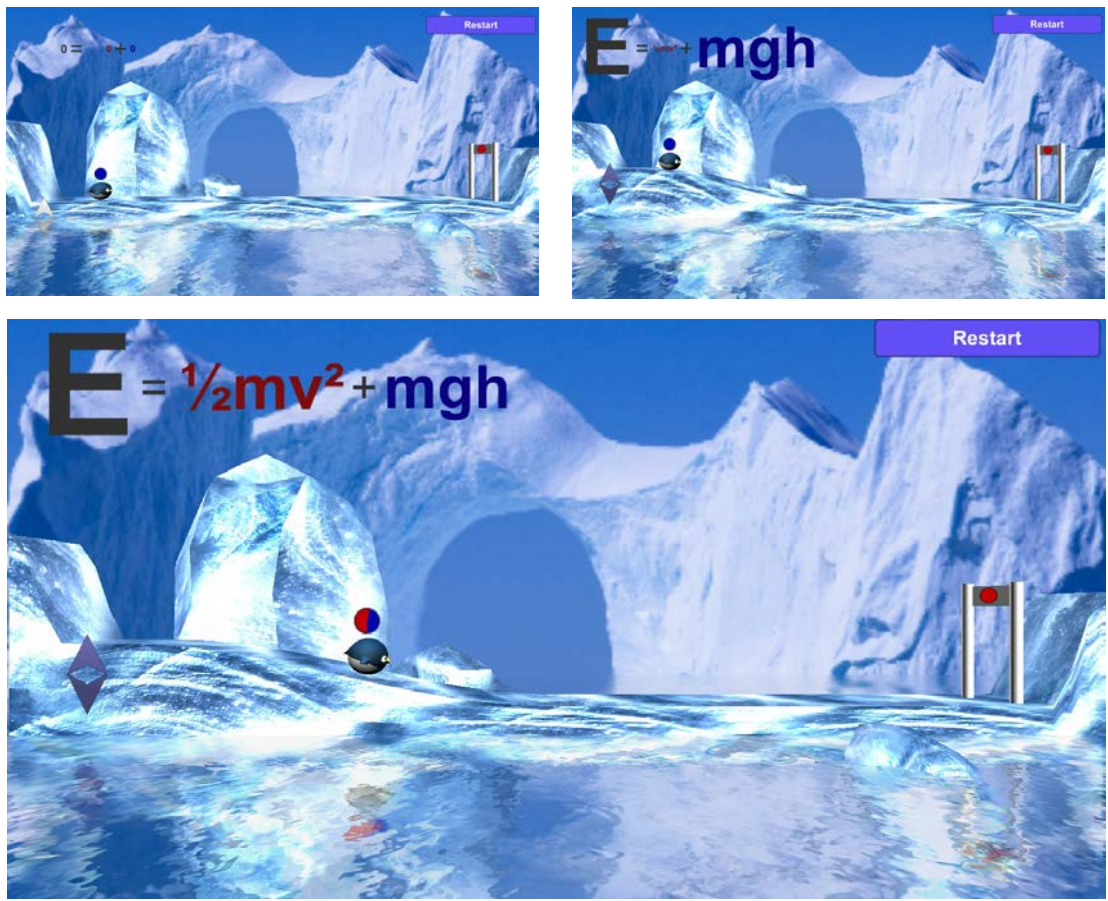

**Figure 1:** Screenshots of the first and simplest level in Potential Penguin. The player can change the height of the hill and thereby the potential energy of the penguin. In the upper-left corner the mathematical expressions for the kinetic ($\frac{1}{2}mv^2$) and potential ($mgh$) energy are displayed. Above the penguin is a pie chart displaying the relative amounts of kinetic (red) and potential (blue) energy. The goal post at the end indicates the maximum relative amount of allowed kinetic energy which in this case is 100%. There is therefore no restriction on the kinetic energy in this particular level.

Through a series of different levels with increasing difficulty the students should form an intuition about the game mechanics, and be able to relate them to the physics concepts as seen in other games (Lieberoth, Pedersen, and Sherson, 2015). In order to generate an explicit connection (Lieberoth and Hansen, 2011; Moon, 2004) to the physics concepts the mathematical expressions are visibly displayed at the top of the screen. The font sizes of the expressions are dynamically changing in order to create a visual connection between the intuitive understanding of the game play and the formula the student would have to work with in order to perform the same calculations outside of the game.

## 3. Proposal for a controlled experiment on learning effects

In order to test the two research questions, we propose a study design comprising of three groups (Table 1). Both of the first two groups play a version of Potential Penguin – one group with changing font sizes for the mathematic expressions, and one group with fixed font sizes.

To test the main hypothesis, a pre-post-test consisting of problem solving exercises with kinetic and potential energy would assess learning gains. An image labelling test where students are asked to identify the different game elements in physics terms would also be administered to both Potential Penguin playing groups to reveal any differences in the vocabulary the students apply in their understanding of the physics problems. If a stronger connection has been established by the changing font sizes we would expect the first group's vocabulary to contain more usage of physics terms compared to game terms (as per Lieberoth, Pedersen & Sherson, 2015).

Our design setup allows us to explore several supporting variables, so data will be gathered from Potential Penguin while the students play to record time needed to solve the levels, mouse movements and attempts needed. Exploring these concurrent data streams for a sufficient N of students could reveal to which degree the students have mastered and understood the game mechanics.

To test the second hypothesis Group 3 will act as active controls, playing other casual games from our website, that similar to Potential Penguin require some amount of puzzle solving and fine motor skill, but which do not require any physics knowledge. This would ensure that the subjects in the control group are equally challenged on similar parameters as Potential Penguin.

Afterwards all groups would play Quantum Moves. A new image label test could reveal a transfer of terms applied in Potential Penguin especially when comparing to a third control group who have only played Quantum Moves. Finally, the ability to understand and pick-up the game mechanics of Quantum Moves will be investigated with data collected from the game and with the results from a questionnaire including the competence and controls scales from the Player Experience of Need Satisfaction (PENS) survey (Rigby and Ryan, 2007). A subset of participants in each potential penguin playing group could also be investigated using eye tracking, in order to assess how much visual attention is given to the mathematical expressions in each condition, and to what extent this measure of fixation duration correlates with game performance on one hand and explicit physics understanding on the other.

**Table 1:** Design for controlled test of Potential Penguin. Comparisons between Group 1 and 2 will investigate the effect of the adjusting equation sizes. Comparisons with Group 3 could reveal the extent of knowledge transfer from Potential Penguin to Quantum Moves.

| Group 1: | Group 2 | Group 3 |
|---|---|---|
| Potentials image labels | Potentials image labels | |
| Problem solving | Problem solving | |
| Play Potential Penguin with changing font sizes | Play Potential Penguin with fixed font sizes | Play unrelated game |
| Potentials image labels | Potentials image labels | |
| Problem solving | Problem solving | |
| Play Quantum Moves | Play Quantum Moves | Play Quantum Moves |
| PENS: Competence and Controls | PENS: Competence and Controls | PENS: Competence and Controls |
| Quantum image labels | Quantum image labels | Quantum image labels |

## 4. Conclusion

How to effectively build in learning goals into the core mechanics of a game, such that explicit connections between the game and the learning content are formed, is a matter of ongoing investigation within the field of game based learning. With the study of Potential Penguin outlined here, we aim to answer part of the question by testing the effect including of dynamic visuals of the learning content.

With this study we are also beginning to test to which degree knowledge of both the game mechanics and the underlying concepts can be transferred from one game to another: From classical physics to quantum physics, and real citizen science problem solving.

**Acknowledgements**

The authors acknowledge the Lundbeck Foundation and Aarhus University Research Foundation (AU Ideas) for financial support.